\documentstyle[12pt]{article}
\begin{document}
\ifx\TwoupWrites\UnDeFiNeD\else\target{\magstepminus1}{11.3in}{8.27in}
	\source{\magstep0}{7.5in}{11.69in}\fi
\newfont{\fourteencp}{cmcsc10 scaled\magstep2}
\newfont{\titlefont}{cmbx10 scaled\magstep2}
\newfont{\authorfont}{cmcsc10 scaled\magstep1}
\newfont{\fourteenmib}{cmmib10 scaled\magstep2}
	\skewchar\fourteenmib='177
\newfont{\elevenmib}{cmmib10 scaled\magstephalf}
	\skewchar\elevenmib='177
\newif\ifpUbblock  \pUbblocktrue
\newcommand\nopubblock{\pUbblockfalse}
\newcommand\topspace{\hrule height 0pt depth 0pt \vskip}
\newcommand\pUbblock{\begingroup \tabskip=\hsize minus \hsize
	\baselineskip=1.5\ht\strutbox \topspace-2\baselineskip
	\halign to\hsize{\strut ##\hfil\tabskip=0pt\crcr
	\the\Pubnum\crcr\the\date\crcr}\endgroup}
\newcommand\YITPmark{\hbox{\fourteenmib YITP\hskip0.2cm
        \elevenmib Uji\hskip0.15cm Research\hskip0.15cm Center\hfill}}
\renewcommand\titlepage{\ifx\TwoupWrites\UnDeFiNeD\null\vspace{-1.7cm}\fi
	\YITPmark
\vskip0.6cm
	\ifpUbblock\pUbblock \else\hrule height 0pt \relax \fi}
\newtoks\date
\newtoks\Pubnum
\newtoks\pubnum
\Pubnum={YITP/U-\the\pubnum}
\date={\today}
\newcommand{\frontpageskip}{\vspace{12pt plus .5fil minus 2pt}}
\renewcommand{\title}[1]{\frontpageskip
	\begin{center}{\titlefont #1}\end{center}\par}
\renewcommand{\author}[1]{\frontpageskip\par\begin{center}
	{\authorfont #1}\end{center}
	\nobreak
	}
\newcommand{\andauthor}{\frontpageskip\centerline{and}\author}
\newcommand{\authors}{\frontpageskip\noindent}
\newcommand{\address}[1]{\par\begin{center}{\sl #1}\end{center}\par}
\newcommand{\andaddress}{\par\centerline{\sl and}\address}
\renewcommand{\thanks}[1]{\footnote{#1}}
\renewcommand{\abstract}{\par\frontpageskip\centerline{\fourteencp Abstract}
	\vspace{8pt plus 3pt minus 3pt}}
\newcommand\YITP{\address{Uji Research Center,
	       Yukawa Institute for Theoretical Physics\\
               Kyoto University,~Uji 611,~Japan\\}}
\thispagestyle{empty}
%
\pubnum{95-25}
\date{June, 1995}
\titlepage

\baselineskip 0.6cm
\title{\Large\sc Density Fluctuations in Inflationary Models\\
with Multiple Scalar Fields}

\author{\sc Jun'ichi Yokoyama}
\YITP

\abstract{Making use of the primordially isocurvature fluctuations,
which are generated in inflationary models with multiple scalar
fields, we make a phenomenological model that predicts formation of
primordial black holes which can account for the massive compact halo
objects recently observed.
}
\hyphenation{non-re-nor-mal-iza-tion}
\newcommand{\gsim}{\mbox{\raisebox{-1.0ex}{$~\stackrel{\textstyle >}
{\textstyle \sim}~$ }}}
\newcommand{\lsim}{\mbox{\raisebox{-1.0ex}{$~\stackrel{\textstyle <}
{\textstyle \sim}~$ }}}
\newcommand{\eV}{\mbox{eV}}
\newcommand{\MeV}{\mbox{MeV}}
\newcommand{\GeV}{\mbox{GeV}}
\newcommand{\TeV}{\mbox{TeV}}
\newcommand{\second}{\mbox{sec}}
\newcommand{\phione}{\phi_1}
\newcommand{\phitwo}{\phi_2}
\newcommand{\phithree}{\phi_3}
\newcommand{\phij}{\phi_j}
\newcommand{\phionec}{\phi_{1\rm c}}
\newcommand{\phitwol}{\phi_{2l}}
\newcommand{\phitwof}{\phi_{2\rm f}}
\newcommand{\tauc}{\tau_{\rm c}}
\newcommand{\taul}{\tau_{l}}
\newcommand{\tauk}{\tau_{\rm k}}
\newcommand{\taus}{\tau_{\ast}}
\newcommand{\taum}{\tau_{\rm m}}
\newcommand{\rmacho}{r_{\rm m}}
\newcommand{\delphione}{\delta\phi_1}
\newcommand{\delphitwo}{\delta\phi_2}
\newcommand{\delphitwof}{\delta\phi_{2\rm f}}
\newcommand{\delphithree}{\delta\phi_3}
\newcommand{\delphij}{\delta\phi_j}
\newcommand{\delphii}{\delta\phi_i}
\newcommand{\phitwom}{\phi_{2\rm m}}
\newcommand{\phithreem}{\phi_{3\rm m}}
\newcommand{\phitwomin}{\phi_{2\rm m}}
\newcommand{\phitwok}{\phi_{2\bf k}}
\newcommand{\lambdaone}{\lambda_1}
\newcommand{\lambdatwo}{\lambda_2}
\newcommand{\lambdathree}{\lambda_3}
\newcommand{\calr}{{\cal R}}
\newcommand{\bx}{{\bf x}}
\newcommand{\by}{{\bf y}}
\newcommand{\br}{{\bf r}}
\newcommand{\bfk}{{\bf k}}
\newcommand{\bkp}{{\bf k'}}
\newcommand{\khor}{k_{\rm hor}}
\newcommand{\beq}{\begin{equation}}
\newcommand{\eeq}{\end{equation}}
\newcommand{\beqa}{\begin{eqnarray}}
\newcommand{\eeqa}{\end{eqnarray}}
\newcommand{\order}{\cal O}
\newcommand{\mpl}{M_{Pl}}
\newcommand{\lmk}{\left(}
\newcommand{\rmk}{\right)}
\newcommand{\lkk}{\left[}
\newcommand{\rkk}{\right]}
\newcommand{\lnk}{\left\{}
\newcommand{\rnk}{\right\}}
\newcommand{\msolar}{M_\odot}
\newcommand{\omegabh}{\Omega_{\rm BH}}
\newcommand{\deltabar}{\overline{\delta}}
\newcommand{\deltabarbh}{\overline{\delta}_{\rm BH}}
\newcommand{\pa}{\Phi_A}
\newcommand{\ph}{\Phi_H}
\newcommand{\rhotot}{\rho_{\rm tot}}
\newpage

In inflationary universe models \cite{inf} with multiple scalar fields,
not only adiabatic \cite{fluc} but also primordially isocurvature
fluctuations \cite{iso} are generated in general.
For example, if we consider
ordinary slow roll-over inflation models in Brans-Dicke gravity
\cite{BD}, the Brans-Dicke dilaton acquires isocurvature fluctuation
out of quantum fluctuation in the inflationary regime.  In this
particular model, however, the isocurvature fluctuation does not play
any important role in structure formation because the energy density
of the Brans-Dicke field remains as small as $\sim \omega^{-1}$ times
the total energy density, where $\omega \gsim 500$ is the Brans-Dicke
parameter \cite{SY}.

On the other hand, primordially isocurvature fluctuations can be
cosmologically important if energy density of its carrier becomes
significant in a later epoch \cite{iso}.
Furthermore it is relatively easier to imprint a nontrivial feature on
the spectral shape of the isocurvature fluctuations.  Making use of
this property here we construct a model which possesses a peak in the
spectrum of total density fluctuation at the horizon crossing.  Then
we can produce a significant amount of primordial black holes (PBHs)
around the horizon mass scale when the fluctuation at the horizon
crossing becomes maximal.  To be specific, we choose the model
parameters so that one can account for the observed massive compact
halo objects (MACHOs) \cite{macho} in terms of PBHs thus produced.
That is, our goal here is to produce PBHs of mass $\sim 0.1\msolar$ with
the abundance of $\sim 20\%$ of the halo mass, corresponding to about
$10^{-3}$ times the critical density \cite{machoproc}.

PBHs are formed if initial density fluctuations grow sufficiently and
a high density region collapses within its gravitational radius.
First let us review its formation process.
The background spacetime of the early universe dominated by radiation
is satisfactorily described by
the spatially-flat Friedmann universe,
\beq
ds^2= -dt^2+a^2(t)\lkk dr^2+r^2\lmk d\theta^2+\sin^2\theta d\varphi^2\rmk\rkk,
\eeq
whose expansion rate is given by
\beq
H^2(t)\equiv \lmk\frac{\dot{a}}{a}\rmk^2=\frac{8\pi G}{3}\rho(t),
\eeq
with $\rho(t)$ being the background energy density and a dot denotes
time derivation.  Following Carr
\cite{Carr}, let us consider a spherically symmetric high density
region with its initial radius, $R(t_0)$, larger than the horizon
scale $\sim t_0$.  The assumption of spherical symmetry will be
justified below.  Then the perturbed region locally constitutes a
spatially closed Friedmann universe with a metric,
\beq
ds^2=-dt'^2+R^2(t')\lkk \frac{dr^2}{1-\kappa r^2}
+r^2(d\theta^2+\sin^2\theta d\varphi^2)\rkk,~~~~~\kappa >0,
\eeq
Then the Einstein equation reads
\beq
H^{'2}(t')\equiv \lmk\frac{1}{R}\frac{dR}{dt'}\rmk^2
= \frac{8\pi G}{3}\rho_+(t')-\frac{\kappa}{R^2(t')},
\eeq
there, where $\rho_+$ is the local energy density.
One can choose the coordinate so that both the background and the
perturbed region have the same expansion rate initially at
$t=t'=t_0$.  Then the initial density contrast, $\delta_0$,
satisfies
\beq
\delta_0 \equiv \frac{\rho_{+0}-\rho_0}{\rho_0}
=\frac{\kappa}{H_0^2R_0^2},
\eeq
where a subscript $0$ implies values at $t_0$.
It can be shown that the two time variables are related by \cite{Harrison}
\beq
  (1+\delta_0)^{\frac{3}{4}}\frac{dt'}{R(t')}=\frac{dt}{a(t)}.
\eeq

The perturbed region will eventually stop expanding at $R \equiv R_c$, which
is obtained from
\beq
0=\frac{8\pi G}{3}\rho_0(1+\delta_0)\lmk\frac{R_0}{R_c}\rmk^4
-\frac{\kappa}{R_c^2}=H_0^2(1+\delta_0)\lmk\frac{R_0}{R_c}\rmk^4
-\frac{R_0^2H_0^2}{R_c^2}\delta_0,
\eeq
as
\beq
R_c=\sqrt{\frac{1+\delta_0}{\delta_0}}R_0 \simeq
\delta_0^{-\frac{1}{2}}R_0,
\eeq
corresponding to the epoch
\beq
t_c \simeq \frac{t_0}{\delta_0}.
\eeq
The perturbed region must be larger than the Jeans scale, $R_J$,  in
order to contract further against the pressure gradient, while it
should be smaller than the horizon scale to avoid formation of a
separate universe.
We thus require
\beq
  R_J \simeq c_st_c \lsim R_c \lsim t_c,
\eeq
or
\beq
  c_s \lsim \frac{R_c}{t_c} \simeq
\frac{R_0}{t_0}\delta_0^{\frac{1}{2}} \lsim 1,
\eeq
where $c_s$ is the sound velocity equal to $1/\sqrt{3}$ in the
radiation dominated era.
Since $R_0\delta_0^{1/2}/t_0$ is time independent, it suffices to
calculate the constraint on $\delta$ at a specific epoch, say, when
the region enters the Hubble radius, $2R=2t$.  We find that the
amplitude should lie in the range
\beq
  \frac{1}{3}\lsim \delta(R=t) \lsim 1.
\eeq

The gravitational radius, $R_g$, of the perturbed region in the beginning of
contraction with $R_c \simeq c_st_c$ is given by
\beq
  R_g=2GM \simeq H^2R_c^3 \simeq \frac{R_c^3}{t_c^2} \simeq c_s^2R_c
\lsim R_c.
\eeq
This is somewhat smaller than $R_c$ but it also implies that a black
hole will be formed soon after the high density region starts
contraction.
Thus we expect that a black hole with a mass around a horizon mass at
$t=t_c$ will result and it has in fact been shown by numerical
calculations \cite{numerical} that the final mass of the black hole is
about $\order\rm(10^{\pm 0.5})$ times the horizon mass at that time.
It has been discussed that these black holes do not accrete
surrounding matter very much and that their mass do not increase even
one order of magnitude \cite{Carr}.  Note also that evaporation due to
the Hawking radiation is unimportant for $M \gg 10^{15}$g.

Since the horizon mass at the time $t$ is given by
\beq
  M_{\rm hor}= 10^5\lmk \frac{t}{1 \rm sec}\rmk\msolar,
\eeq
what is required in order to produce a significant number
of PBHs with mass $M\sim 0.1\msolar$
is the sufficient amplitude of density fluctuations on
the horizon scale at $t \sim 10^{-6}$sec.
Because the initial mass fraction of PBHs, $\beta$, is related with the
present fraction $\omegabh$ as
\beq
 \beta=\frac{a(10^{-6}\second)}{a(t_{\rm eq})}\omegabh
  \cong 10^{-8}\omegabh,
\eeq
where $t_{\rm eq}$ is the equality time, only an extremely tiny fraction
of the universe, $\beta \simeq 10^{-11}$ should collapse into black
holes.

That is, the probability of having a density contrast $1/3 \lsim \delta
\lsim 1$ on the horizon scale at $t\sim 10^{-6}$sec should be equal
to $\beta$.
Let us assume  density fluctuations on the relevant scale
obey the Gaussian statistics with the dispersion $\deltabarbh \ll 1$,
which is the correct property for the model introduced here.
Then the probability of PBH formation is estimated as
\beqa
\beta &=& \int_{1/3}^1 \frac{1}{\sqrt{2\pi}\,\deltabarbh}
\exp\lmk -\frac{\delta^2}{2\deltabarbh^2}\rmk d\delta  \nonumber \\
&\simeq& \int_{1/3}^{1/3+\order{\rm(\deltabarbh^2)}}
 \frac{1}{\sqrt{2\pi}\,\deltabarbh}
\exp\lmk -\frac{\delta^2}{2\deltabarbh^2}\rmk d\delta \nonumber \\
&\simeq& \deltabarbh\exp\lmk -\frac{1}{18\deltabarbh^2}\rmk,  \label{F}
\eeqa
which implies that we should have
\beq
\deltabarbh \simeq 0.05,
\eeq
to produce appropriate amount of PBHs.
Although it is true that an exponential accuracy is required on the
amplitude of fluctuations in order to produce a desired amount of PBHs
we have not been able to obtain the correspondence between $\deltabarbh$
and $\omegabh$ with such an accuracy because numerical coefficients
appearing in the above expressions, such as 18 in (\ref{F}), have been
calculated based on a rather qualitative discussion.
We therefore will not attempt exceedingly quantitative analysis in
what follows.

Note also that for $\deltabarbh =0.05$ the threshold of PBH
formation, $\delta=1/3$, corresponds to 6.4 standard deviation.
It has been argued by Doroshkevich \cite{Doro}
that such a high peak has very likely
a spherically symmetric shape.  Thus the assumption of spherical
symmetry in the above discussion is justified and it is also expected that
gravitational wave produced during PBH formation is negligibly small.

Since the primordial amplitude of density perturbations on large
scales probed by the anisotropy of the background radiation \cite{cobe}
is known to be $\deltabar \simeq 10^{-5}$, the primordial fluctuations
must have such a spectral shape that it has an amplitude of $10^{-5}$
on large scales, sharply increases by a factor of $10^4$ on the mass
scale of PBHs, and decreases again on smaller scales at the time of
horizon crossing.  It is difficult
to produce such a spectrum of fluctuations in inflationary cosmology
with a single component.

In generic inflationary models with a single scalar field $\phi$, which
drives inflation with a potential $V[\phi]$, the root-mean-square
amplitude of adiabatic fluctuations generated is given by
\beq
\frac{\delta\rho}{\rho}(r) \equiv \deltabar\lmk r(\phi)\rmk
\cong \frac{8\sqrt{6\pi}V[\phi]^{\frac{3}{2}}}{V'[\phi]\mpl^3},
\label{single}
\eeq
on the comoving scale $r(\phi)$ when that scale reenters the Hubble
radius \cite{fluc}.
The right-hand-side is evaluated when the same scale leaves
the horizon during inflation.  Because of the slow variation of $\phi$
and rapid cosmic expansion during inflation, (\ref{single}) implies an
almost scale-invariant spectrum in general.
Nonetheless one could in principle obtain various shapes of
fluctuation spectra making use of the nontrivial dependence of
$\deltabar \lmk r(\phi)\rmk$ on $V[\phi]$ \cite{Hodges}.
In order to obtain a desirable spectrum for PBH formation with a
mountain on a particular scale, we must employ a scalar potential with
two breaks and a plateau in between \cite{Ivanov}.  Such a solution is not
aesthetically appealing.

Here we instead consider an inflation model with multiple scalar fields
in which not only adiabatic but also primordially isocurvature
fluctuations are produced.  In fact it is much easier to imprint
nontrivial structure on the isocurvature spectrum as mentioned in the
beginning.

We introduce three scalar fields $\phione$, $\phitwo$, and $\phithree$
in order to generate the desired spectrum of density
fluctuation. $\phione$ is the inflaton field which induces the new
inflation \cite{newinf}
with a double-well potential, starting its evolution near
the origin where its potential is approximated as
\beq
U[\phione ] \cong V_0 -\frac{\lambdaone}{4}\phione^4, \label{infpot}
\eeq
so that the Hubble parameter during inflation, $H_I$, is given by
\[  H_I^2=\frac{8\pi V_0}{3\mpl^2}. \]
On the other hand, $\phitwo$ is a long-lived scalar field which
induces primordially isocurvature fluctuations that contribute to
black hole formation later.  Finally $\phithree$ is an auxiliary field
coupled to both $\phione$ and $\phitwo$ and it changes the effective
mass of the latter to imprint a specific feature on the spectrum of
its initial fluctuations.

We adopt the following model Lagrangian.
\beq
{\cal L}=\frac{1}{2}(\partial \phione)^2 +\frac{1}{2}(\partial
\phitwo)^2 +\frac{1}{2}(\partial \phithree)^2
-V[\phione, \phitwo, \phithree] +{\cal L}_{\rm int},
\eeq
where ${\cal L}_{\rm int}$ represents interaction of $\phij$'s
with other fields.
Here $V[\phione,\phitwo,\phithree]$ is the effective
scalar potential which is a
polynomial of $\phij$'s only up to the fourth order except for the
inflaton sector $U[\phione]$:
\beq
V[\phione,\phitwo,\phithree]= U[\phione]+\frac{\epsilon}{2}
(\phione^2-\phionec^2)\phithree^3 + \frac{\lambdathree}{4}\phithree^4
-\frac{\nu}{2}\phithree^2\phitwo^2 +\frac{\lambdatwo}{4}\phitwo^4
+\frac{1}{2}m_2^2\phitwo^2,   \label{totpot}
\eeq
where $\lambda_j$, $\epsilon$, $\nu$, $\phionec$, and $m_2$ are
positive constants.  $m_2$ is assumed to be much smaller than the
scale of inflation, $H_I$, and it does not affect the dynamics of
$\phitwo$ during inflation.  Hence we ignore it for the moment.

Let us briefly outline how the system evolves before presenting its
detailed description.
In the early inflationary stage $\phione$ is smaller than $\phionec$,
and $\phithree$ has its potential minimum off the origin.  Then
$\phitwo$ also settles down to a nontrivial minimum, where it can have
an effective mass larger than $H_I$ so that its quantum fluctuation is
suppressed.  As $\phione$ becomes larger than $\phionec$, $\phithree$
rolls down to its origin.  Then the potential of $\phitwo$ also
becomes convex and its amplitude gradually decreases
due to its quartic term.  However, since its
potential is now nearly flat, its motion is extremely slow with its
effective mass smaller than $H_I$ well until the end of inflation.  In
this stage quantum fluctuations are generated to $\phitwo$  with a
nearly scale-invariant spectrum.  Thus the initial spectrum of the
isocurvature fluctuations has a scale-invariant spectrum with a
cut-off on a large scale.

After inflation, the Hubble parameter starts to decrease in the
reheating processes.  As it becomes smaller than the effective
mass of $\phitwo$, the latter starts rapid coherent oscillation.
$\phitwo$ dissipates its energy in the same way as radiation in the
beginning when its oscillation is governed by the quartic term.  But
later on when $\lambdatwo\phitwo^2$ becomes smaller than $m_2^2$, its
energy density decreases more slowly in the same manner as
nonrelativistic matter.  Thus $\phitwo$ contributes to the total
energy density more and more later, which implies that the total
density fluctuation due to the primordially isocurvature fluctuations
or $\phitwo$ grows with time.  Since what is relevant for PBH
formation is the magnitude of fluctuations at the horizon crossing, we
thus obtain a spectrum with a larger amplitude on larger scale until
the cut-off scale in the initial spectrum is reached, that is, it has
a single peak on the mass scale of PBH formation.

In the above scenario we have assumed that $\phitwo$ survives until
after the PBH formation.  On the other hand, were $\phitwo$ stable, it
would soon dominate the total energy density of the universe in
conflict with the successful nucleosynthesis.  As a natural
possibility we assume that $\phitwo$ decays through gravitational
interaction, so that it does not leave any unwanted relics with its
only trace being the tiny amount of PBHs produced.

Having stated the outline of the scenario, we now proceed to its
detailed description and
obtain constraints on the model parameters to produce
the right amount of PBHs on the right scale.
First we consider the evolution of the homogeneous part of the fields.
During inflation, the evolution of the inflaton is governed by the
$U[\phione]$ part of the potential.  Solving the equation of motion
with the slow-roll approximation,
\beq
  3H_I\dot{\phione}\cong -U'[\phione] \cong \lambdaone\phione^3,
\eeq
we find
\beq
  \lambdaone\phione^2(t)=\frac{\lambdaone\phi_{1i}^2}
  {1+\frac{2\lambdaone\phi_{1i}^2}{3H_I^2}H(t-t_i)},  \label{infsol}
\eeq
where $\phi_{1i}$ is the field amplitude at some initial epoch $t_i$.
The above approximate solution remains valid until $|U''[\phione]|$
becomes as large as $9H_I^2$ at $t\equiv t_f$, when inflationary
expansion is terminated and we find
$\phione^2(t_f)=3H_I^2/\lambdaone$.  Then (\ref{infsol}) can also be
written as
\beq
  \lambdaone\phione^2(t)=\frac{3H_I^2}{2H_I(t_f-t)+1}
  \equiv \frac{3H_I^2}{2\tau(t)+1} \cong \frac{3H_I^2}{2\tau(t)},
\eeq
where $\tau(t)$ is the $e$-folding number of exponential expansion
after $t\ (< t_f)$ and the last approximation is valid when $\tau \gg 1$.
{}From now on, we
often use $\tau(t)$ as a new time variable or to refer to the comoving
scale leaving the Hubble radius at $t$.
Note that it is a decreasing function of $t$.

As stated above, we are taking a view that $\phione$
determines fate of $\phithree$ and that $\phithree$ controls evolution of
$\phitwo$ but not vice versa.  In order that $\phithree$ does not
affect evolution of $\phione$ the inequality
\beq
  \lambdaone\lambdathree \gg \epsilon^2,  \label{L13}
\eeq
must be satisfied, while we must have
\beq
  \lambdatwo\lambdathree \gg \nu^2,  \label{L23}
\eeq
so that $\phitwo$ does not affect the motion of $\phithree$.  We assume
these inequalities hold below.

When $\phione < \phionec$, both $\phitwo$ and $\phithree$ have
nontrivial minima which we denote by $\phitwom$ and $\phithreem$,
respectively. From
\beq
  V_2=-\nu\phithree^2\phitwo +\lambdatwo\phitwo^3 =0,
\eeq
and
\beq
  V_3=\epsilon (\phione^2-\phionec^2)\phithree
  +\lambdathree\phithree^3 -\nu\phitwo^2\phithree =0,
\eeq
with $V_j \equiv \partial V/\partial \phij$, we find
\beqa
  \lambdatwo\phitwom^2(t) &=& \nu\phithreem^2(t)  \\
  \lambdathree\phithreem^2(t)&=& \epsilon(\phionec^2-\phione^2(t))
  +\nu\phitwom^2(t) \cong \epsilon(\phionec^2-\phione^2(t)),
\eeqa
where (\ref{L23}) was used in the last expression.  In the early
inflationary stage when $\phione \ll \phionec$, the effective
mass-squared of $\phij$, $V_{jj}$, at the potential minimum
is given by
\beq
  V_{22}[\phione,\phitwom,\phithreem]=2\lambdatwo\phitwom^2
  =\frac{2\nu\epsilon}{\lambdathree}(\phionec^2-\phione^2)
  \simeq \frac{2\nu\epsilon}{\lambdathree}\phionec^2
  =\frac{3\nu\epsilon}{\lambdaone\lambdathree\tauc}H_I^2,
\eeq
\beq
  V_{33}[\phione,\phitwom,\phithreem]
  =\frac{3\epsilon}{\lambdaone\tauc}H_I^2,
\eeq
where $\tauc$ is the epoch when $\phione=\phionec$.
We choose parameters such that
\beq
  \nu\epsilon > \lambdaone\lambdathree\tauc,~~~{\rm and}~~~
  \epsilon > \lambdaone\tauc.   \label{twoineq}
\eeq
Then $V_{22}$ and $V_{33}$ are larger than $H_I^2$ initially at the
potential minimum, so that
both $\phitwo$ and $\phithree$ settle down to $\phitwom(t)$ and
$\phithreem(t)$, respectively.

$\phithreem(t)$ decreases down to zero at $\tau=\tauc$ when
$V_{33}[\phithreem]$ also vanishes.  Then $V_{33}[\phithreem=0]$
starts to increase according to $\phione$
and soon acquires a large positive value, which implies that
$\phithree$ practically traces the evolution of $\phithreem(t)$ down to
zero without delay.  On the other hand, $\phitwo$ evolves somewhat
differently because it does not acquire a positive effective mass from
$\phithree$ at
the origin.  Although $\phitwo(t)$ traces $\phitwom(t)$ initially, as
$V_{22}[\phitwom]$ becomes smaller it can no longer catch up with
$\phitwom(t)$.  From a generic property of a scaler field with a small
mass in the De Sitter background, one can show that this happens when
the inequality
\beq
  \left| \frac{1}{\phitwom(t)}\frac{d\phitwom(t)}{dt}\right| >
  \frac{V_{22}[\phitwom]}{3H},
\eeq
gets satisfied, or at
\beq
  \tau=\lmk 1+
  \sqrt{\frac{\lambdaone\lambdathree}{2\nu\epsilon}}\rmk\tauc
  \equiv \taul \cong \tauc,
\eeq
with
\beq
  \phitwo^2(\taul)\equiv \phitwol^2 =
  \sqrt{\frac{\nu\epsilon}{2\lambdaone\lambdathree}}
   \frac{3H_I^2}{2\lambdatwo\taul}.  \label{phitwoel}
\eeq
Thus $\phitwo$ slows down its evolution.  In the
meantime $\phithree$ vanishes.  Then $\phitwo$ is governed by the
quartic potential.
We can therefore summarize its evolution during inflation as
\beqa
\phitwo^2(\tau)\cong \lnk
  \begin{array}{@{\,}ll@{\,}}
     \phitwom^2(\tau)=\frac{3\nu\epsilon H_I^2}
     {2\lambdaone\lambdatwo\lambdathree}
     \lmk\frac{1}{\tauc}-\frac{1}{\tau}\rmk,  & \tau \gsim \taul, \\
     \phitwol^2\lkk 1+\frac{2\lambdatwo\phitwol^2}{3H_I^2}
     (\taul-\tau)\rkk^{-1},
     & 0< \tau \lsim \taul.   \rule{0mm}{6ex}
  \end{array}
\right.  \label{phitwosinka}
\eeqa
Since $\lambdatwo\phitwol^2$ is adequately smaller than $H_I^2$,
$\phitwo$ remains practically constant in the latter regime.  Let us
also write down time dependence of its effective mass-squared for
later use.
\beqa
V_{22}[\phitwo]\cong \lnk
  \begin{array}{@{\,}ll@{\,}}
     \frac{\nu\epsilon H_I^2}
     {\lambdaone\lambdathree}
     \lmk\frac{1}{\tauc}-\frac{1}{\tau}\rmk,  & \tau \gsim \taul,
 \\
     3\lambdatwo\phitwol^2\lkk 1+\frac{2\lambdatwo\phitwol^2}{3H_I^2}
     (\taul-\tau)\rkk^{-1},
     & 0< \tau \lsim \taul.   \rule{0mm}{6ex}
  \end{array}
\right.
\eeqa

Next we consider fluctuations in both the scalar fields and metric
variables in a consistent manner. We adopt Bardeen's
gauge-invariant variables $\pa$ and $\ph$ \cite{Bardeen},
with which the perturbed
metric can be written as
\beq
  ds^2=-(1+2\pa)dt^2+a(t)^2(1+2\ph)d\bx ^2,
\eeq
 in the longitudinal gauge.
In this gauge scalar field fluctuation $\delphij$ coincides with
the corresponding gauge-invariant variable by itself.

 Assuming an $\exp(i\bfk\bx)$ spatial
dependence and working in the Fourier space, the perturbed Einstein
and scaler field equations are given by
\beqa
\pa+\ph &=& 0\\
\dot{\ph}+H\ph &=& - 4\pi G
(\dot{\phione}\delphione +\dot{\phitwo}\delphitwo
+\dot{\phithree}\delphithree) \label{ad}
\eeqa
\beqa
\ddot{\delphij}+3H\dot{\delphij}
+\lmk \frac{k^2}{a^2(t)} + V_{jj}\rmk \delphij
=2V_j\pa+\dot{\pa}\dot{\phij}
-3\dot{\ph}\dot{\phij}-\sum_{i\neq j}V_{ji}\delta\phi_i, \label{delphieq}
\eeqa
Note that all the fluctuation variables are functions of $\bfk$ and
$t$.

The above system is quite complicated at a glance.  However, using
constraints on various model parameters we have obtained so far, {\it
i.e.} (\ref{L13}), (\ref{L23}), and (\ref{twoineq}), it can somewhat
be simplified.  First, since $\phithree$ has an effective mass larger
than $H_I^2$ during inflation except in the vicinity of $\tau=\tauc$,
quantum fluctuation on $\delphithree$ is suppressed and moreover
its energy density practically vanishes by the end of inflation.  We can
therefore neglect fluctuations in $\phithree$.  On the other hand, we
can show that
\beq
|\dot{\phitwo}| \sim
\sqrt{\frac{\nu\epsilon}{\lambdatwo\lambdathree}} |\dot{\phione}| \ll
|\dot{\phione}|,
\eeq
with the help of (\ref{L23}) and (\ref{twoineq}).  Hence (\ref{ad})
and (\ref{delphieq}) with $j=1$
reduce to
\beq
\dot{\ph}+H\ph = - 4\pi G\dot{\phione}\delphione  \label{ad1}
\eeq
\beqa
\ddot{\delphione}+3H\dot{\delphione}
+\lmk \frac{k^2}{a^2(t)} + V_{11}\rmk \delphione  \label{delphione}
=2V_1\pa-4\dot{\ph}\dot{\phione}.
\eeqa
Thus only $\phione$ contributes to adiabatic fluctuations and it can
be calculated in the same manner as in the new inflation model with a
single scalar field.  This is as expected because $\phione$ dominates
the energy density during inflation.
In fact since we are only interested in the growing mode on the
super-horizon regime which turns
out to be weakly time-dependent as can be seen from the final result,  we
can consistently neglect time derivatives of metric perturbations and
terms with two time derivatives in (\ref{ad1}) and (\ref{delphione})
during inflation.  We thus find
\beq
  \ph =-\pa \cong -\frac{4\pi G}{H_I}\dot{\phione}\delphione.
\eeq
The resultant amplitude of
scale-invariant adiabatic fluctuations depends on $\lambdaone$ and one
can normalize its value using the COBE observation \cite{cobe}as
\beq
  \lambdaone=1.3\times10^{-13}.  \label{L1}
\eeq

On the other hand, $\delphitwo$ satisfies (\ref{delphieq}) with
$j=2$.  From quantum field theory in De Sitter spacetime, it has an RMS
amplitude $\delphitwo \cong (H^2/2k^3)^{1/2}$ when the $k$-mode leaves
the Hubble radius if $V_{22}$ is not too large.
Since $V_2$ vanishes when $\phitwo=\phitwom$ and $V_2\pa$
remains small even for $\tau \leq \taul$, we can neglect all the terms
in the right-hand side, to yield
\beq
  \ddot{\delphitwo}+3H_I\dot{\delphitwo}+ V_{22}\delphitwo \cong 0,
  \label{delphitwo}
\eeq
when $k \ll a(t)H_I$.
We can find a WKB solution with the appropriate initial condition,
\beqa
  \delphitwo (k,t) &\cong& \sqrt{\frac{H_I^2}{2k^3}}
  \lmk\frac{S(t_k)}{S(t)}\rmk^{\frac{1}{2}}\exp\lnk\int^t_{t_k}
  \lkk S(t')H_I-\frac{3}{2}H_I\rkk dt'\rnk,   \label{WKB}  \\
  S(t)&\equiv& \frac{3}{2}\sqrt{ 1-\frac{4V_{22}}{9H^2}},
   \rule{0mm}{6ex} \nonumber
\eeqa
where $t_k$ is the time when $k$-mode leaves the Hubble radius:
$k=a(t_k)H_I$.  The above expression is valid when $|\dot{S}|\ll
S^2$.  In terms of $\tau$,
(\ref{WKB}) can be expressed as
\beq
  \delphitwo (k,\tau) \cong \sqrt{\frac{H_I^2}{2k^3}}
  \lmk\frac{S(\tauk)}{S(\tau)}\rmk^{\frac{1}{2}}
  \exp\lnk\int^{\tauk}_{\tau}S(\tau')d\tau'
  -\frac{3}{2}(\tauk -\tau)\rnk,  \label{delphitwosol}
\eeq
\[
  S(\tau)=\lnk
  \begin{array}{@{\,}ll@{\,}}
     \frac{3}{2}
     \lmk 1-\frac{\taus}{\tauc}+\frac{\taus}{\tau}\rmk^{\frac{1}{2}},
     & \tau \gsim \taul, \\
     \frac{3}{2}
     \lmk 1-\frac{2\lambdatwo\phitwol^2}{3H_I^2}\lkk 1+
     \frac{2\lambdatwo\phitwol^2}{3H_I^2}(\taul-\tau)\rkk^{-1}
     \rmk^{\frac{1}{2}},
     & 0< \tau \lsim \taul,   \rule{0mm}{11mm}
  \end{array}
\right.
\]
where $\tauk\equiv \tau(t_k)$ and $\taus\equiv
\frac{4\nu\epsilon}{3\lambdaone\lambdathree}$.  The above equality is
valid until the end of inflation at $t=t_f$ or $\tau=0$.

Let us assume the universe is rapidly and efficiently reheated at
$t=t_f$ for simplicity to avoid further complexity. (see, {\it e.g.}
\cite{reheating}
for recent discussion on efficient reheating.)  Then the reheat
temperature is given by
\beq
  T_R \cong 0.1\sqrt{H_I\mpl}.
\eeq
If there is no further significant entropy production later, one can
calculate the epoch, $\tau(L)$, when the comoving length scale
corresponding to $L$ pc today left the Hubble horizon during inflation
as
\beq
  \tau(L)=37+\ln\lmk\frac{L}{1\, \rm pc}\rmk +
  \frac{1}{2}\ln\lmk\frac{H_I}{10^{10}\ \rm GeV}\rmk.  \label{scale}
\eeq
Then the comoving horizon scale at
$t=10^{-6}$sec, or $L=0.03$ pc corresponds to $\tau\cong34\equiv\taum$ and
the present horizon scale $\simeq 3000$Mpc to $\tau\cong 59$.

On the other hand, $\phitwo(t)$ and $\delphitwo(\bfk,t)$ evolve according
to
\beqa
 \ddot{\phitwo} +3H\dot{\phitwo}+\lambdatwo\phitwo^3 + m_2^2\phitwo
 &=&0,  \label{phitwoeq} \\
 \ddot{\delphitwo}+3H\dot{\delphitwo}+(3\lambdatwo\phitwo^2+m_2^2)
 \delphitwo &\cong& 0,  \label{resonanteq}
\eeqa
where the Hubble parameter is now time-dependent: $H=1/2t$, and the
latter equation is valid for $k \ll aH$.

When $H^2$ becomes smaller than $\lambdatwo\phitwo^2~(\gg m_2^2)$,
both $\phitwo$ and $\delphitwo$ start rapid oscillation around the
origin.  Using (\ref{delphitwosol}) one can express the
amplitude of the gauge-invariant comoving fractional density
perturbation of $\phitwo$, $\Delta_2$, as
\beq
   \Delta_2 = \frac{1}{\rho_2}\lmk\dot{\phitwo}\dot{\delphitwo} -
   \ddot{\phitwo}\delphitwo -\dot{\phitwo}^2\pa\rmk \simeq
   \left.4\frac{\delphitwo}{\phitwo}\right|_{\tau=0}
	\equiv 4\frac{\delphitwof}{\phitwof},
\eeq
in the beginning of oscillation.  Here
\beq
   \rho_2 \equiv \frac{1}{2}\dot{\phitwo}^2
    +\frac{\lambdatwo}{4}\phitwo^4 + \frac{1}{2}m_2^2\phitwo^2
\eeq
is the energy density of $\phitwo$.

Using the virial theorem one can easily show that it decreases
in proportion to
$a^{-4}(t)$ as long as $\lambdatwo\phitwo^2 \gsim m_2^2$.  Thus the
amplitude of $\phitwo$ decreases with $a^{-1}(t)$.  On the other hand,
$\delphitwo$ has a rapidly oscillating mass term when
$\lambdatwo\phitwo^2 \gsim m_2^2$, which causes parametric amplification.
We have numerically solved equations (\ref{phitwoeq}) and
(\ref{resonanteq}) with various initial conditions with $|\phitwo| \gg
|\delphitwo|$ initially.  We have found that in all cases the
amplitude of $\delphitwo$ remains constant as long as $m_2$ is negligible.
Thus $\Delta_2$ increases in proportion to $t^{1/2}$ in this regime
and $\Delta_2$ becomes as large as
\beq
  \Delta_2 \cong 4\frac{\delphitwof}{\phitwof}
  \lmk\frac{\lambdatwo\phitwof^2}{m_2^2}\rmk^{\frac{1}{2}}, \label{deltatwo}
\eeq
while the ratio of $\rho_2$ to the total energy density, $\rhotot$,
which is now dominated by radiation, remains constant:
\beq
  \frac{\rho_2}{\rhotot}=2\pi\lmk\frac{\phitwof}{\mpl}\rmk^2.
\eeq

As $\lambdatwo\phitwo^2$ becomes smaller than $m_2^2$, $\phitwo$ and
$\delphitwo$ come to satisfy the same equation of motion, see
(\ref{phitwoeq}) and (\ref{resonanteq}), and $\Delta_2$ saturates to
the constant value (\ref{deltatwo}).  At the same time $\rho_2$ starts
to decrease less rapidly than radiation, in proportion to $a^{-3}(t)$.
Since $\Delta_2$ contributes to the total comoving density fluctuation
by the amplitude
\beq
   \Delta \cong \frac{\rho_2}{\rhotot}\Delta_2,
\eeq
it increases in proportion to $a(t) \propto t^{1/2}$.
In the beginning of this stage, we find $H^2 \cong
m_2^4/\lambdatwo\phitwof^2$, to yield
\beq
\Delta \cong 8\pi\frac{\delphitwof}{\phitwof}
  \lmk\frac{\lambdatwo\phitwof^2}{m_2^2}\rmk^{\frac{1}{2}}
  \lmk\frac{\phitwof}{\mpl}\rmk^2
  \lmk\frac{2m_2^2 t}{\sqrt{\lambdatwo}\phitwof}\rmk^{\frac{1}{2}},
\eeq
at a later time $t$.

In order to relate it with the initial condition required for PBH
formation, we must estimate it at the time $k$-mode reenters the
Hubble radius, $t_k^\ast$, defined by
\beq
  k=2\pi a(t_k^\ast)H(t_k^\ast)=
  \frac{\pi a_f}{\lmk t_k^\ast t_f\rmk^{\frac{1}{2}}}.
\eeq
Since $k$ can also be expressed as $k=2\pi a_f e^{-\tauk}H_I$, the
amplitude of comoving density fluctuation at $t=t_k^\ast$ is given by
\beq
 \Delta(k,t_k^\ast) \cong 8\pi\frac{\delphitwof}{\phitwof}
  \lmk\frac{\sqrt{\lambdatwo}\phitwof}{H_I}\rmk^{\frac{1}{2}}
  \lmk\frac{\phitwof}{\mpl}\rmk^2 e^{\tauk}.
\eeq

In the beginning of this article
 we discussed the necessary condition on the amplitude of
fluctuations for PBH formation using the uniform Hubble constant
gauge.  Hence we should calculate the predicted amplitude in this
gauge, which is a linear combination of $\Delta$ and the
gauge-invariant velocity perturbation.  However, in the present case
in which $\Delta$ grows in proportion to $a(t)$ in the
radiation-dominant universe, one finds that the latter quantity
vanishes and that density fluctuation in the uniform Hubble constant
gauge coincides with $\Delta$.  Thus we finally obtain the quantity to
be compared with $\deltabarbh$ in (\ref{F}), namely, the
root-mean-square amplitude of density fluctuation on scale $r=2\pi/k$
at the horizon crossing, $\deltabar (r)$, as
\beq
  \deltabar (r) = \lkk\frac{4\pi k^3}{(2\pi)^3}
   |\Delta (k,t_k^\ast)|^2\rkk^{\frac{1}{2}}
 \cong 4\lmk\frac{\sqrt{\lambdatwo}H_I\phitwof^3}{\mpl^4}\rmk^{\frac{1}{2}}
   e^{\tauk}C_f(\tauk,\tauc,\taus),
\eeq
with
\beq
    C_f(\tauk,\tauc,\taus)\equiv\lmk\frac{S(\tauk)}{S(0)}\rmk^{\frac{1}{2}}
    \exp\lnk \int^{\tauk}_0 S(\tau')d\tau' -\frac{3}{2}\tauk\rnk,
\eeq
where we have used (\ref{delphitwosol}).
We also find
\beq
   \phitwof^2= \sqrt{\frac{3\taus}{8}}
   \lmk 1+\sqrt{\frac{2}{3\taus}}\rmk^{-1}
   \lmk 1+\sqrt{\frac{3\taus}{8}}\rmk^{-1}
   \frac{3H_I^2}{2\lambdatwo\tauc},
\eeq
from (\ref{phitwoel}) and (\ref{phitwosinka}).

The remaining task is to choose values of parameters so that
$\deltabar (r)$ has a peak on the comoving horizon scale at
$t=10^{-6}$sec, which we denote by $\rmacho$, corresponding to
$\tauk=\taum \cong 34$, with its amplitude $\deltabar (\rmacho)\cong 0.05$.
We thus require
\beqa
\frac{d\ln \deltabar(r)}{d\tauk} &=&
\frac{S'(\tauk)}{2S(\tauk)}+S(\tauk)-\frac{1}{2} \nonumber
\\
&=&-\frac{1}{4\tauk}\frac{\taus\tauc}
{\lkk\tauc\tauk+\taus\lmk \tauc-\tauk \rmk\rkk}
+\frac{3}{2}\lmk
1-\frac{\taus}{\tauc}+\frac{\taus}{\tauk}\rmk^{\frac{1}{2}}
-\frac{1}{2}
\eeqa
vanishes at $\tauk=\taum$, which gives us a relation between $\tauc$
and $\taus$.  Since $\tauc$ roughly corresponds to the comoving scale
where scale-invariance of primordial fluctuation $\Delta_2$ is broken,
the peak at $\deltabar (\rmacho)$ becomes the sharper, the closer $\tauc$
approaches $\taum$.

For example, if we take $\tauc=30$ we find
\beq
  \taus=\frac{4\nu\epsilon}{3\lambdaone\lambdatwo}=200,  \label{taustars}
\eeq
so that
\beq
  C_f=0.13~~~{\rm and}~~~\phitwof^2=0.045\frac{H_I^2}{\lambdatwo}.
\eeq
In order to have $\deltabar (\rmacho)=0.05$, we require
\beq
  \frac{1}{\sqrt{\lambdatwo}}
  \lmk\frac{H_I}{\mpl}\rmk^2=1.7\times 10^{-15},  \label{LH}
\eeq
which can easily be satisfied with some reasonable choices of
$\lambdatwo$ and  $H_I$.  However, it is not the final
constraint.  Since we are assuming that the universe is dominated by
radiation at this time, we require
\beq
  \frac{\rho_2}{\rhotot}=2\pi\lmk\frac{\phitwof}{\mpl}\rmk^2
  \lmk\frac{m_2^2}{\sqrt{\lambdatwo}\phitwof H_I}\rmk^{\frac{1}{2}}
  e^{\taum} \ll 1.  \label{rhoratio}
\eeq
Furthermore $\phitwo$ should decay some time after $t=10^{-6}$sec so
as not to dominate the energy density of the universe which would
hamper the primordial nucleosynthesis.  Assuming that it decays only
through gravitational interaction, its life time is given by
\beq
  \tau_{\phitwo} \cong \frac{\mpl^2}{m_2^3}
  = 10^{-5.5}\lmk\frac{m_2}{10^{6.5}\GeV}\rmk^{-3}\sec.
\eeq

Now we have displayed all the necessary equalities and inequalities
the model parameters should satisfy.  Since there is a wide range of
allowed region in the multi-dimensional space of parameters, we do not
work out the details of the constraints but simply give one example of
their values with which all the requirements are satisfied.
\beqa
   H_I &=& 1.7\times 10^{10}\GeV, \nonumber\\
   m_2 &=& 3.2\times 10^6\GeV, \nonumber\\
   \lambdaone &=& 1.3\times 10^{-13}, \\
   \lambdatwo &=& 1.4\times 10^{-6},  \nonumber\\
   \lambdathree &=& \nu = 6.7\times 10^{-8},  \nonumber\\
   \epsilon &=& 2.0\times 10^{-11}, \nonumber
\eeqa
for which $\rho_2/\rhotot=0.1$ at $t=10^{-6}$sec and inequalities
(\ref{L13}) and (\ref{L23}) are maximally satisfied.

Thus we have reached a model with the desired feature making use of a
simple polynomial potential (\ref{totpot}).  In order to set the order of
magnitude of the mass
scale of the black holes and that of their abundance correctly,
we must tune some combinations of model parameters such as
(\ref{taustars}) and (\ref{LH}) with two digits' accuracy. However,
there exists a wide range of allowed region in the parameter space to
realize it.

\vspace{0.8cm}
%


\end{document}